\documentclass[10pt,a4paper]{article}
\usepackage[utf8]{inputenc}
\usepackage[T1]{fontenc}
\usepackage{lmodern}
\usepackage[a4paper,margin=2.5cm]{geometry}
\usepackage{amsmath,amssymb,amsfonts}
\usepackage{graphicx}
\usepackage{textcomp}
\usepackage{xcolor}
\usepackage{float}
\usepackage{caption}
\usepackage[colorlinks=true,linkcolor=blue,citecolor=blue,urlcolor=blue]{hyperref}
\usepackage{xurl}
\usepackage{authblk}
\usepackage{booktabs}
\usepackage{siunitx}
\usepackage[backend=biber, style=ieee, isbn=false, sortcites, maxbibnames=6, minbibnames=1, safeinputenc]{biblatex}
\DeclareSourcemap{
  \maps[datatype=bibtex]{\map{\step[fieldsource=abstract,null]\step[fieldset=abstract,null]}}
}

\title{\textbf{Post-Quantum Entropy as a Service for Embedded Systems}}

\author[1]{Javier Blanco-Romero\thanks{Corresponding author: frblanco@pa.uc3m.es}}
\author[2,3]{Yuri Melissa Garcia-Ni\~no}
\author[1]{Florina Almenares Mendoza}
\author[1]{Daniel D\'iaz-S\'anchez}
\author[1]{Carlos Garc\'ia-Rubio}
\author[1]{Celeste Campo}

\affil[1]{Department of Telematic Engineering, Universidad Carlos III de Madrid, Legan\'es, Madrid, Spain}
\affil[2]{Universidad Carlos III de Madrid, Av. de la Universidad, 30, 28911 Legan\'es, Madrid, Spain}
\affil[3]{Universidad Industrial de Santander, Bucaramanga, Santander, Colombia}

\begin{document}

\maketitle

\begin{abstract}
Embedded cryptography stands or falls on entropy quality, yet small devices have few trustworthy sources and little tolerance for heavyweight protocols. We build a Quantum Entropy as a Service (QEaaS) system that moves QRNG-derived entropy from a Quantis device to ESP32-class clients over post-quantum-secured channels. On the server side, the design exposes two paths: direct quantum entropy through a custom OpenSSL provider and mixed entropy through the Linux system pool. On the client side, we extend libcoap's Zephyr support, integrate wolfSSL-based DTLS 1.3 into the CoAP stack, and add a BLAKE2s entropy pool that preserves the standard Zephyr extraction interface while introducing an injection API for server-provided entropy. Benchmarks on ESP32 hardware, targeting 100 iterations per configuration, show that ML-KEM-512 completes a DTLS 1.3 handshake in 313 ms on average without certificate verification, 35\% faster than ECDHE P-256. Pairing ML-KEM-512 with ML-DSA-44 lowers the mean to 225 ms. Certificate verification adds roughly 194 ms for ECDSA but only 17 ms for ML-DSA-44, so the fully post-quantum configuration remains 63\% faster than classical ECDHE P-256 with ECDSA even under full verification. Local BLAKE2s pool operations stay below 0.1 ms combined. On this platform, post-quantum key exchange and authentication are not only feasible; they are faster than the classical baseline.
\end{abstract}


\section{Introduction}

Random numbers sit underneath nearly every security primitive: key generation, nonces, password salting, sequence numbers. If the entropy source is weak, the rest of the stack inherits that weakness. IoT devices make this problem harder rather than easier. They usually expose a hardware RNG, but the path from physical noise source to software API is often opaque, and vendor post-processing can hide the raw generator behaviour. A practical response is to combine several sources instead of trusting a single one~\cite{donenfeld2022rng}. In that setting, Quantum Random Number Generators (QRNGs), which draw randomness from irreducible quantum processes~\cite{schmidt1970quantum}, are attractive external inputs. The catch is transport: once entropy leaves the QRNG host, the distribution channel must itself survive the arrival of quantum-capable attackers. Post-quantum cryptography (PQC) addresses that problem through quantum-resistant algorithms~\cite{bernstein2025post}. NIST has now standardized the lattice-based ML-KEM for key encapsulation~\cite{bos2018crystals,NIST2024fips203} and ML-DSA for digital signatures~\cite{NIST2024fips204}.

This paper studies that full path end to end. We build a Quantum Entropy as a Service (QEaaS) architecture that delivers QRNG-backed entropy to constrained devices over CoAP protected with post-quantum DTLS. Unlike earlier EaaS systems built around HTTP or HTTPS, our design is shaped around the constraints of embedded clients: CoAP transport, explicit local pooling, and an ESP32 implementation that can actually run the stack.

\section{Background and Related Work}
\label{sec:back_relatedwork}

Cryptographic systems need high-quality entropy to produce unpredictable keys. Vassilev et al.~\cite{vassilev2016entropy} outline Entropy-as-a-Service (EaaS) as a way to supply entropy from distributed sources, which motivates remote entropy distribution for cloud and IoT settings.

Several systems already show that remote quantum entropy is viable. Huang et al.~\cite{huang2021quantum} built a cloud platform on Alibaba Cloud that combines four QRNG types and deployed it for production services such as Alipay. Kozlovi{\v{c}}s et al.~\cite{kozlovivcs2022poster} proposed a remote QRNG service with quantum-safe links, X.509 certificates carrying post-quantum algorithms, and bidirectional WebSocket connections. Kumar et al.~\cite{kumar2022quantum} generated quantum random numbers on IBM's cloud quantum platform with Hadamard-gate circuits and validated the output with NIST SP~800-90B~\cite{turan2018sp800_90b} and SP~800-22 tests. Blanco et al.~\cite{blanco2025integration} studied QRNG integration with post-quantum cryptography in TLS workflows.

Taken together, these results show that remote quantum entropy is plausible, but they mostly target conventional computing environments. In parallel, PQC benchmarking on constrained hardware has moved from standalone kernels~\cite{kannwischer2019pqm4} to complete protocol stacks. B\"urstinghaus-Steinbach et al.~\cite{burstinghaus2020post} integrate Kyber and SPHINCS+ into mbed TLS on ESP32 and show that lattice-based key exchange can compete with ECC. Segatz and Al Hafiz~\cite{segatz2025efficient} then push Kyber further on ESP32 with dual-core parallelism. Nielsen et al.~\cite{nielsen2025post} show that post-quantum signatures are also feasible on LoRa ESP32 platforms across several NIST security levels. At the transport level, Sikeridis et al.~\cite{sikeridis2020post} and Sosnowski et al.~\cite{sosnowski2023performance} measure PQC handshake latency in TLS~1.3 under realistic network conditions, and Tasopoulos et al.~\cite{tasopoulos2023energy} extend that analysis to energy consumption on STM32 with wolfSSL. What is still missing is the intersection of these lines of work: remote entropy delivery for embedded systems, protected by post-quantum transport, over a protocol that constrained devices actually prefer. The local entropy-pool question is also largely open in this setting.

Entropy pool design grounds our client-side architecture. Dodis et al.~\cite{dodis2013security} formalized robustness requirements for PRNGs with entropy accumulation and showed that the Linux /dev/random pool fell short of this model. Coretti et al.~\cite{coretti2019seedless} strengthened the framework by proving that cryptographic hash functions serve as seedless extractors, and the Linux kernel adopted this result by replacing its SHA-1/LFSR-based pool with BLAKE2s~\cite{aumasson2013blake2,donenfeld2022rng}, gaining better input mixing, deterministic output, and replacing the earlier LFSR-based mixing stage. Chung et al.~\cite{chung2024provable} prove that Linux-DRBG achieves a 128-bit security bound in the seedless robustness model for the analyzed Linux 6.4.8 construction.

\section{System Architecture}
\label{sec:architecture}

\begin{figure}[H]
  \centering
  \includegraphics[width=1\linewidth]{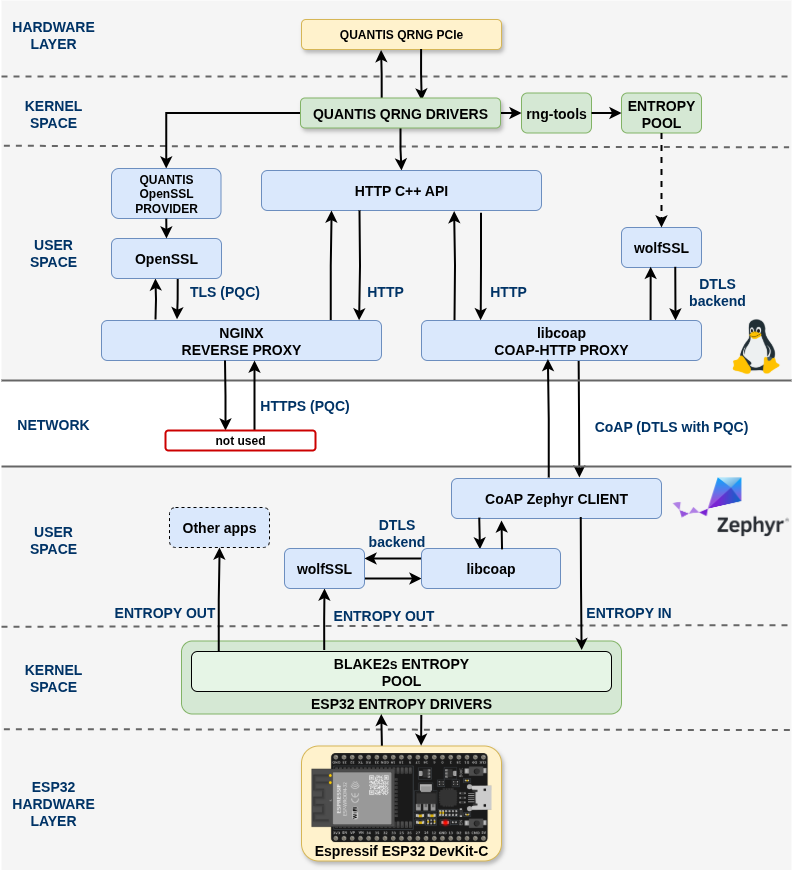}
  \caption{QEaaS system architecture for quantum entropy distribution from Quantis QRNG via dual access (OpenSSL and Linux entropy pool) to ESP32 clients with BLAKE2s entropy pools, using post-quantum cryptography for HTTPS and CoAP.}
  \label{fig:system_architecture}
\end{figure}

Our system moves quantum-generated entropy from a centralized hardware source to embedded IoT devices. The architecture spans the whole path, from QRNG hardware to transport security to the local pool on the microcontroller, as shown in Figure~\ref{fig:system_architecture}.

A Quantis QRNG PCIe-240M~\cite{IDQuantiquePCIe240M} produces the raw quantum entropy. Linux kernel drivers expose it to user space, where our implementation offers two access paths: direct access through a custom OpenSSL provider, and mixed access through the Linux entropy pool after rng-tools folds the QRNG stream together with other system sources.

The service exposes both HTTPS and secure CoAP, with post-quantum cryptography protecting each path. A CoAP-HTTP proxy compliant with RFC 7252~\cite{rfc7252} forwards constrained-client requests carrying a Proxy-URI option to the HTTP backend. ESP32 clients running Zephyr RTOS use that path and mix the received bytes into a local BLAKE2s entropy pool. The server infrastructure~\cite{qeaas_server} runs in Docker containers with automated setup scripts.

\subsection{Server Architecture}

The server architecture centers on a Quantis QRNG PCIe-240M generating quantum entropy, exposed through \texttt{/dev/qrandom0} by its Linux kernel drivers. The PCIe-240M uses photon counting on an image sensor illuminated by a light-emitting diode; after post-processing via Universal-2 hash functions the output delivers \SI{58}{\mega\bit\per\second} of random data compliant with NIST~SP~800-90A/B/C and SP~800-22~\cite{IDQuantiquePCIe240M}. The rng-tools daemon can be configured to read from this device and inject quantum entropy into the system entropy pool. The first approach provides direct quantum entropy through a custom OpenSSL provider~\cite{quantis_qrng_openssl_integration}, building on our previous work exploring QRNG integration methods in OpenSSL~\cite{blanco2024evaluating}. The second uses the Linux entropy pool where quantum entropy mixes with other system sources via rng-tools and is then consumed by applications linked against the standard Linux randomness interfaces.

An Nginx reverse proxy uses the direct OpenSSL quantum provider for HTTPS clients, while a CoAP-HTTP proxy~\cite{coap2http_proxy2023} uses wolfSSL as its DTLS backend and forwards constrained-client requests to the HTTP backend through libcurl. The HTTP API listens on port~6065 and exposes a \texttt{/random\_number/\{num\_bytes\}} endpoint that currently validates requests in the 1 to 256 bytes range and returns the result as JSON. In the checked-in implementation this API reads from \texttt{/dev/urandom}, so the proxy/API path is accurately described as consuming the mixed Linux entropy pool rather than directly exporting \texttt{/dev/qrandom0}. This builds upon our previous work on integrating post-quantum cryptography into CoAP protocols via wolfSSL~\cite{blanco2024integrating}.

\subsection{Client Architecture}

The client implementation~\cite{qeaas_esp32_client} targets ESP32 microcontrollers running Zephyr RTOS. Two changes made this practical: extending libcoap's Zephyr support through POSIX API integration~\cite{libcoap_zephyr_pr} and adding a native wolfSSL DTLS backend~\cite{libcoap_wolfssl_pr}. Both were merged upstream into libcoap.

The client uses a split TLS stack. ESP32 WiFi still depends on mbedTLS for connectivity, while CoAP traffic runs over wolfSSL with native ML-KEM (FIPS~203) key encapsulation and ML-DSA-44 (FIPS~204) digital signatures. The integration relies on wolfSSL's OpenSSL compatibility layer (\texttt{OPENSSL\_EXTRA}) for the libcoap DTLS backend and on Zephyr's POSIX API for socket and threading abstractions.

\subsubsection{ESP32 Entropy Pool Implementation}

The standard Zephyr ESP32 entropy driver provides only basic hardware entropy access without pooling mechanisms. In our Zephyr fork~\cite{zephyr_entropy_pool_fork} we extend the entropy driver interface with an optional \texttt{add\_entropy} callback and the public \texttt{entropy\_add\_entropy()} API, allowing external entropy sources to be mixed into a driver's internal pool with an explicit entropy-bit estimate. On top of this interface we implement a BLAKE2s-based pool device inspired by the Linux~5.17+ random subsystem, while preserving the standard \texttt{entropy\_get\_entropy()} extraction API for applications.

The client repository includes automated build scripts that integrate our custom entropy pool into Zephyr on-the-fly during compilation. Our ESP32 implementation maintains 512 bytes of mixed entropy with a 128-byte threshold and wraps the ESP32 TRNG as a backend hardware source selected through devicetree. The BLAKE2s pool automatically refills itself from that backend via \texttt{entropy\_get\_entropy()}, while applications can inject external entropy, such as bytes obtained from the QEaaS server, via \texttt{entropy\_add\_entropy(dev, data, len, entropy\_bits)}. Entropy mixing uses BLAKE2s to combine externally supplied bytes with the current pool state, while the backend hardware source continues to contribute local entropy through asynchronous refill operations.

At the application level, the Zephyr fork exposes a simple model. Applications read local entropy through \texttt{entropy\_get\_entropy()} and can mix in external bytes, including material fetched from the QEaaS service, through \texttt{entropy\_add\_entropy()}. Our benchmark client sends CoAP requests to the proxy with a Proxy-URI targeting the HTTP backend; dedicated microbenchmarks characterize the pool separately.

\section{Experimental Evaluation}
\label{sec:evaluation}

\subsection{Experimental Setup}

The benchmark scenario evaluates the complete QEaaS pipeline in a local LAN configuration where client and server communicate over a WiFi link on the same network, isolating the system from Internet variability.

\subsubsection{Client Hardware and Software}
The client is an \textbf{ESP32-DevKitC V4} board (Espressif, ESP32-WROOM-32 module) featuring a dual-core Xtensa LX6 processor at \SI{240}{\mega\hertz}, \SI{520}{\kilo\byte} SRAM, \SI{4}{\mega\byte} flash, and integrated WiFi~802.11\,b/g/n plus Bluetooth~4.2. The firmware runs Zephyr RTOS with the following software stack:
\begin{itemize}
    \item \textbf{libcoap~4.3.5}~\cite{libcoap_fork} (zephyr\_wolfssl branch) as CoAP client library, communicating via the Proxy-URI option (RFC~7252 Section~5.10.2);
    \item \textbf{wolfSSL~5.8.2} as the DTLS/TLS backend, compiled with support for ML-KEM (FIPS~203) key encapsulation and ML-DSA-44 (FIPS~204) digital signatures using wolfSSL's native implementations, DTLS~1.3, and the \texttt{TLS\_AES\_128\_GCM\_SHA256} cipher suite. A parametric build selects the key-exchange algorithm (ECDHE~P-256, X25519, ML-KEM-512) and signature scheme (ECDSA or ML-DSA-44) at compile time via preprocessor switches. RSA, DH, AES-256, SHA-384, and SHA-512 are disabled to minimise DRAM usage. The wolfSSL memory allocator is redirected from newlib's \texttt{malloc()} (limited to ${\sim}$\SI{22}{\kilo\byte} on ESP32's sbrk heap) to Zephyr's \texttt{k\_malloc()} backed by a \SI{105}{\kilo\byte} system heap, resolving out-of-memory errors during DTLS handshake buffer allocation;
    \item \textbf{BLAKE2s entropy pool} as a custom Zephyr entropy driver (512-byte pool, 128-byte refill threshold) replacing the default ESP32 hardware TRNG driver while wrapping it as a seed source;
    \item \textbf{mbedTLS} (built-in) exclusively for ESP32 WiFi WPA2 connectivity, not used for application-layer cryptography.
\end{itemize}

Table~\ref{tab:memory} summarises the per-variant firmware footprint across all seven algorithm groups (verify and non-verify variants share identical static footprints; only peak heap differs by 0.2 to 1.3\,KiB). Static DRAM is constant at \SI{187}{\kilo\byte} (97.6\% of \SI{192}{\kilo\byte}) for all DTLS variants, as the difference lies only in code and dynamic allocations. Flash ranges from \SI{855}{\kilo\byte} (plain CoAP) to \SI{888}{\kilo\byte} (ML-KEM-512~+~ML-DSA-44). ML-DSA-44 adds 6 to 14\,\si{\kilo\byte} of Code~ROM (depending on shared lattice code with the key exchange) and ${\sim}$\SI{28}{\kilo\byte} of peak heap over the corresponding ECDSA variant. The Zephyr system heap provides \SI{105}{\kilo\byte} for dynamic allocations; peak usage ranges from \SI{34}{\kilo\byte} (plain CoAP) to \SI{97}{\kilo\byte} (ML-KEM-512~+~ML-DSA-44 with verification), leaving \SI{8}{\kilo\byte} of headroom in the most demanding configuration.

\begin{table}[H]
\centering
\caption{ESP32 firmware footprint per algorithm variant. Peak~Heap is the runtime high-water mark across both verify and non-verify variants ($\Delta = 0.2$ to $1.3$\,KiB).}\label{tab:memory}
\begin{tabular}{@{}l S[table-format=3.0] S[table-format=3.0] S[table-format=3.0] S[table-format=2.0]@{}}
\toprule
\textbf{Variant} & {\textbf{Flash}} & {\textbf{Code ROM}} & {\textbf{DRAM}} & {\textbf{Peak Heap}} \\
 & {(\si{\kilo\byte})} & {(\si{\kilo\byte})} & {(\si{\kilo\byte})} & {(\si{\kilo\byte})} \\
\midrule
Plain CoAP (no DTLS)        & 855 & 535 & 187 & 34 \\
\midrule
ECDHE P-256 + ECDSA         & 857 & 537 & 187 & 66 \\
X25519 + ECDSA              & 860 & 540 & 187 & 66 \\
ML-KEM-512 + ECDSA          & 882 & 562 & 187 & 70 \\
\midrule
ECDHE P-256 + ML-DSA-44     & 871 & 551 & 187 & 94 \\
X25519 + ML-DSA-44          & 874 & 554 & 187 & 93 \\
ML-KEM-512 + ML-DSA-44      & 888 & 568 & 187 & 97 \\
\bottomrule
\end{tabular}
\end{table}

\subsubsection{Server Infrastructure}
The server runs on a Gigabyte B250M-DS3H (Intel Core i7-7700, 4 cores/8 threads at up to \SI{4.2}{\giga\hertz}, \SI{32}{\giga\byte} RAM, Ubuntu~22.04) equipped with the Quantis QRNG PCIe-240M card. The ESP32 is placed at ${\sim}$\SI{1}{\metre} from a TP-Link Archer~AX73 router and connects via WiFi~802.11\,n (\SI{2.4}{\giga\hertz}); the router connects to the server's Gigabit Ethernet port. The plain CoAP baseline (\SI{17.6}{\milli\second} mean, Table~\ref{tab:dtls_bench}) characterises the end-to-end WiFi round-trip latency of this link. The server hosts three Docker containers with host networking:
\begin{itemize}
    \item \textbf{QRNG API} (\texttt{qrng-api}): HTTP REST service on port~6065 exposing the \texttt{/random\_number/\allowbreak\{num\_bytes\}} endpoint; the checked-in implementation reads from \texttt{/dev/urandom};
    \item \textbf{CoAP-HTTP proxy}~\cite{coap2http_proxy2023} (\texttt{libcoap-proxy}): RFC~7252-compliant proxy built on libcoap with wolfSSL~5.8.2 (native ML-KEM, no liboqs dependency), listening on CoAP port~5683 and CoAPS port~5684, forwarding Proxy-URI requests to the HTTP API via libcurl;
    \item \textbf{Nginx reverse proxy} (\texttt{oqs-nginx}): post-quantum HTTPS endpoint using OpenSSL with OQS provider.
\end{itemize}

Table~\ref{tab:versions} summarises the software component versions shared across client and server.

\begin{table}[H]
\centering
\caption{Software component versions.}\label{tab:versions}
\begin{tabular}{@{}lll@{}}
\toprule
\textbf{Component} & \textbf{Client (ESP32)} & \textbf{Server (Linux)} \\
\midrule
Zephyr RTOS      & 4.1.0         & ---           \\
wolfSSL           & 5.8.2-stable  & 5.8.2-stable  \\
libcoap           & 4.3.5         & 4.3.5   \\
ML-KEM backend    & native (wolfSSL) & native (wolfSSL) \\
\bottomrule
\end{tabular}
\end{table}

\subsubsection{Transport Configuration}
All benchmarks use \textbf{CoAPS over DTLS~1.3}~\cite{rfc9147} (port~5684) with three key-exchange algorithms (ECDHE~P-256, X25519, ML-KEM-512) paired with two signature schemes (ECDSA and ML-DSA-44). We test each algorithm combination in two certificate verification modes: without verification (\texttt{WOLFSSL\_NO\_VERIFY}), which retains the mandatory CertificateVerify signature exchange but skips certificate chain validation, and with full certificate chain verification enabled, which additionally validates the CA signature on the server certificate. Together with a plain CoAP baseline this yields thirteen configurations. ML-KEM-512 key shares exceed the DTLS MTU, so a ClientHello carrying one would need fragmentation. wolfSSL's fragmentation path (\texttt{WOLFSSL\_DTLS\_CH\_FRAG}) pulls in server-side code incompatible with client-only builds (\texttt{NO\_WOLFSSL\_SERVER}). Instead, the firmware calls \texttt{wolfSSL\_NoKeyShares()} to send an empty first ClientHello, triggering the standard DTLS~1.3 HelloRetryRequest with a stateless cookie (RFC~9147~\S5.1~\cite{rfc9147}). The client then sends a complete second ClientHello carrying the chosen key share. This adds one round trip ($\sim$\SI{18}{\milli\second}) but preserves full DTLS~1.3 handshake semantics with no cryptographic weakening, and all reported latencies already include this cost.

The wolfSSL configuration is aggressively optimised for ESP32 memory constraints: I/O buffers are statically capped at the DTLS MTU (\SI{1400}{\byte}) instead of the default \SI{16384}{\byte} via \texttt{STATIC\_CHUNKS\_ONLY}, the OpenSSL compatibility layer is reduced to the minimal subset required by libcoap, and session caching is disabled. All wolfSSL allocations are redirected to Zephyr's \texttt{k\_malloc()} via \texttt{XMALLOC\_OVERRIDE}, providing \SI{105}{\kilo\byte} of heap versus the ${\sim}$\SI{22}{\kilo\byte} available through newlib's sbrk-based allocator.

\subsubsection{Timing Methodology}
We measure latency using the ESP32 hardware cycle counter via \texttt{k\_cycle\_get\_32()} at \SI{240}{\mega\hertz} ($\approx$\SI{4.17}{\nano\second} per tick), converted to microseconds with \texttt{k\_cyc\_to\_us\_floor32()}. The 32-bit counter wraps every ${\sim}$\SI{17.9}{\second}; unsigned subtraction handles wrap-around correctly. We track heap usage via \texttt{sys\_heap\_runtime\_stats\_get()} at five checkpoints: boot, pre-handshake, periodically during handshake iterations (every 10 iterations), post-handshake, and final. We extract static memory sizes (Flash, Code~ROM, DRAM in Table~\ref{tab:memory}) from the ESP-IDF linker memory map; Peak~Heap is the runtime high-water mark of \texttt{max\_allocated\_bytes}. Phase~1 (local pool) runs 100~iterations per buffer size. Phase~2 (end-to-end) targets 100 fresh-session DTLS handshakes per configuration, each creating a new session, completing the handshake and first CoAP request, and tearing down the session with a \SI{1000}{\milli\second} inter-iteration delay for DTLS variants (\SI{100}{\milli\second} for plain CoAP) to allow server-side session cleanup. A subsequent set of $n{=}100$ CoAP round-trip measurements on an established session provides the baseline RTT. The Phase~2 timer spans from session creation (which triggers the DTLS handshake for CoAPS, or UDP binding for plain CoAP) to first CoAP response receipt; local entropy pool operations are excluded from these timers and characterised independently in Phase~1, so the handshake measurements isolate protocol and cryptographic overhead. We build, flash, and benchmark each algorithm independently with a server restart between runs.

\subsubsection{Statistical Reporting}
Tables report mean $\pm$ standard deviation. Derived comparisons propagate uncertainty via first-order Taylor expansion on the standard error of the mean ($\text{SEM} = \sigma / \sqrt{n}$).

\subsection{Local Entropy Pool Performance}

We first isolate the BLAKE2s pool to see what it costs on its own, without network effects. These measurements come from the plain CoAP firmware run. Every build includes the pool benchmark, but the plain CoAP run is the cleanest reference because it adds no DTLS overhead. We execute 100 iterations per buffer size after 5 warmup iterations.

\begin{table}[H]
\centering
\caption{BLAKE2s entropy pool latency, mean $\pm$ std (\si{\micro\second}), on ESP32 at \SI{240}{\mega\hertz} ($n{=}100$).}
\label{tab:pool_latency}
\begin{tabular}{@{}l ccccc@{}}
\toprule
Operation & {16\,B} & {32\,B} & {64\,B} & {128\,B} & {256\,B} \\
\midrule
Extraction  & $34 \pm 15$ & $44 \pm 14$ & $86 \pm 0$ & $132 \pm 0$ & $224 \pm 0$ \\
Injection   & $4 \pm 5$   & $7 \pm 6$   & $13 \pm 0$ & $26 \pm 0$  & $50 \pm 0$  \\
\bottomrule
\end{tabular}
\end{table}

\begin{figure}[H]
  \centering
  \includegraphics[width=0.85\linewidth]{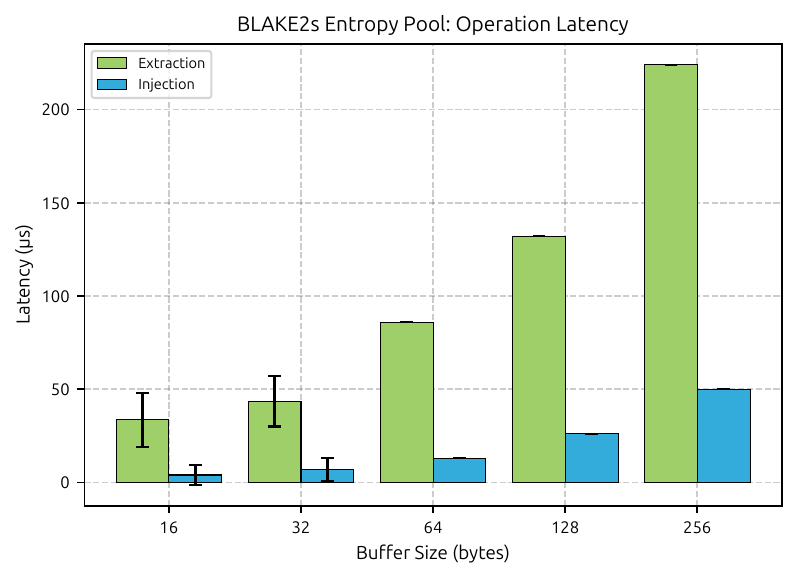}
  \caption{BLAKE2s entropy pool operation latency by buffer size. Extraction requires full pool state hashing while injection performs incremental BLAKE2s mixing.}
  \label{fig:pool_operations}
\end{figure}

Table~\ref{tab:pool_latency} shows a simple pattern. Extraction is slower than injection because extraction hashes the full pool state, whereas injection only mixes new material into the existing state. Both costs grow roughly linearly with buffer size. For buffers of \SI{64}{\byte} and above, the standard deviation drops below \SI{1}{\micro\second}, which means the code path is effectively deterministic at this timer resolution. A 32-byte inject-then-extract cycle costs \SI{51}{\micro\second} in total, so local pool handling is negligible next to any network round trip. Figure~\ref{fig:pool_operations} plots the same trend.

\subsection{End-to-End Distribution Latency}

We then estimate the full entropy-delivery cycle by combining the Phase~1 pool measurements ($n{=}100$) with the Phase~2b CoAP round-trip measurements ($n{=}100$) taken over an already established DTLS session. The modelled cycle is the one the client actually cares about: send a CoAP GET with Proxy-URI, let the proxy query the HTTP entropy API through libcurl, receive the JSON response for \texttt{/random\_number/8}, inject the returned bytes into the BLAKE2s pool, and finally extract entropy locally.

\begin{table}[H]
\centering
\caption{End-to-end QEaaS distribution latency, mean $\pm$ std (\si{\milli\second}), over CoAPS with session reuse (entropy pool: $n{=}100$; CoAP RTT: $n{=}100$ per algorithm, pooled across 12 DTLS configurations).}
\label{tab:full_cycle}
\begin{tabular}{@{}l c@{}}
\toprule
Component & {Mean $\pm$ Std (\si{\milli\second})} \\
\midrule
CoAP round-trip    & $24 \pm 3$    \\
Entropy injection  & ${<}0.1$      \\
Entropy extraction & ${<}0.1$      \\
\midrule
Full cycle         & $24 \pm 3$    \\
\bottomrule
\end{tabular}
\end{table}

Table~\ref{tab:full_cycle} makes the bottleneck obvious. The CoAP round trip dominates at \SI{24.1}{\milli\second} on average, and that number already includes WiFi transmission, proxy handling, the HTTP backend query, and the response path back to the ESP32. Its \SI{3.4}{\milli\second} standard deviation is ordinary WiFi jitter. By contrast, local pool operations add only \SI{0.050}{\milli\second} combined (Table~\ref{tab:pool_latency}), which is too small to matter at this scale.

\subsection{Impact of Algorithm Selection}

Table~\ref{tab:dtls_bench} presents the DTLS~1.3 handshake latency and subsequent CoAP round-trip time for all twelve algorithm combinations: three key exchanges (ECDHE~P-256, X25519, ML-KEM-512) paired with two signature schemes (ECDSA and ML-DSA-44) in two certificate verification modes (disabled and enabled), plus a plain CoAP (no DTLS) baseline. Each algorithm targets 100 independent iterations (warmup iteration excluded); each handshake creates a fresh DTLS session, completes the full handshake and first CoAP request, then tears down the session, isolating per-handshake variability. We measure subsequent RTT over $n{=}100$ requests on an established session and report all values as mean $\pm$ std.

\begin{table}[H]
\centering
\caption{DTLS~1.3 handshake + first CoAP request latency and subsequent CoAP round-trip time per algorithm combination, mean $\pm$ std (target $n{=}100$ iterations per configuration).}\label{tab:dtls_bench}
\begin{tabular}{@{}l l l c c@{}}
\toprule
\textbf{Key Exchange} & \textbf{Signature} & \textbf{Verify} & {\textbf{Handshake + 1\textsuperscript{st} req (\si{\milli\second})}} & {\textbf{RTT (\si{\milli\second})}} \\
\midrule
\multicolumn{3}{@{}l}{Plain CoAP (no DTLS)} & $18 \pm 3$ & $16 \pm 4$ \\
\midrule
ECDHE P-256  & ECDSA     & ---  & $479 \pm 11$  & $23 \pm 2$ \\
ECDHE P-256  & ECDSA     & \checkmark & $668 \pm 10$  & $24 \pm 4$ \\
X25519       & ECDSA     & ---  & $819 \pm 11$  & $24 \pm 2$ \\
X25519       & ECDSA     & \checkmark & $1008 \pm 9$  & $23 \pm 2$ \\
ML-KEM-512   & ECDSA     & ---  & $313 \pm 6$   & $24 \pm 2$ \\
ML-KEM-512   & ECDSA     & \checkmark & $517 \pm 41$  & $25 \pm 5$ \\
\midrule
ECDHE P-256  & ML-DSA-44 & ---  & $405 \pm 64$  & $24 \pm 4$ \\
ECDHE P-256  & ML-DSA-44 & \checkmark & $416 \pm 15$  & $24 \pm 3$ \\
X25519       & ML-DSA-44 & ---  & $745 \pm 27$  & $25 \pm 4$ \\
X25519       & ML-DSA-44 & \checkmark & $759 \pm 19$  & $24 \pm 4$ \\
ML-KEM-512   & ML-DSA-44 & ---  & $225 \pm 9$   & $25 \pm 4$ \\
ML-KEM-512   & ML-DSA-44 & \checkmark & $249 \pm 20$  & $25 \pm 3$ \\
\bottomrule
\end{tabular}
\end{table}

ML-KEM-512 is the fastest key exchange in every configuration we tested: \SI{313}{\milli\second} with ECDSA and no certificate verification, and \SI{225}{\milli\second} with ML-DSA-44 in the same mode. Without verification, it beats ECDHE~P-256 by 35\% and X25519 by 62\% under ECDSA signatures. Replacing ECDSA with ML-DSA-44 reduces latency for every key exchange: ECDHE~P-256 falls from 479 to \SI{405}{\milli\second}, X25519 from 819 to \SI{745}{\milli\second}, and ML-KEM-512 from 313 to \SI{225}{\milli\second}. The reason is narrow and concrete. DTLS~1.3 always includes a CertificateVerify step in which the server signs the handshake transcript and the client checks that signature. On this platform, ML-DSA-44 handles that pair of operations faster than ECDSA.

Turning certificate verification on exposes an even sharper difference between the signature schemes. Averaged across key exchanges, ECDSA verification adds \SI{194.2 \pm 1.6}{\milli\second}, while ML-DSA-44 adds only \SI{16.5 \pm 2.6}{\milli\second}. The same gap appears in every algorithm pair: P-256 adds $+$188.8 vs. $+$11.5~ms, X25519 adds $+$189.4 vs. $+$13.7~ms, and ML-KEM-512 adds $+$204.5 vs. $+$24.3~ms for ECDSA and ML-DSA-44, respectively. On the ESP32, certificate validation with ML-DSA-44 is therefore about 12$\times$ faster than with ECDSA. That is why the fully post-quantum configuration with verification enabled (ML-KEM-512~+~ML-DSA-44, \SI{249}{\milli\second}) still beats the classical verified baseline (ECDHE~P-256~+~ECDSA, \SI{668}{\milli\second}) by 63\%. Without verification, the same post-quantum pair is 53\% faster than the classical baseline. Since all three key exchanges target roughly the same 128-bit classical security level, this is a like-for-like comparison. Once the session exists, the transport becomes nearly boring: all DTLS variants settle to about \SI{24}{\milli\second} RTT, while plain CoAP sits at \SI{15.7}{\milli\second}, so the remaining difference is just symmetric record-layer cost.

\begin{figure}[H]
  \centering
  \includegraphics[width=1\linewidth]{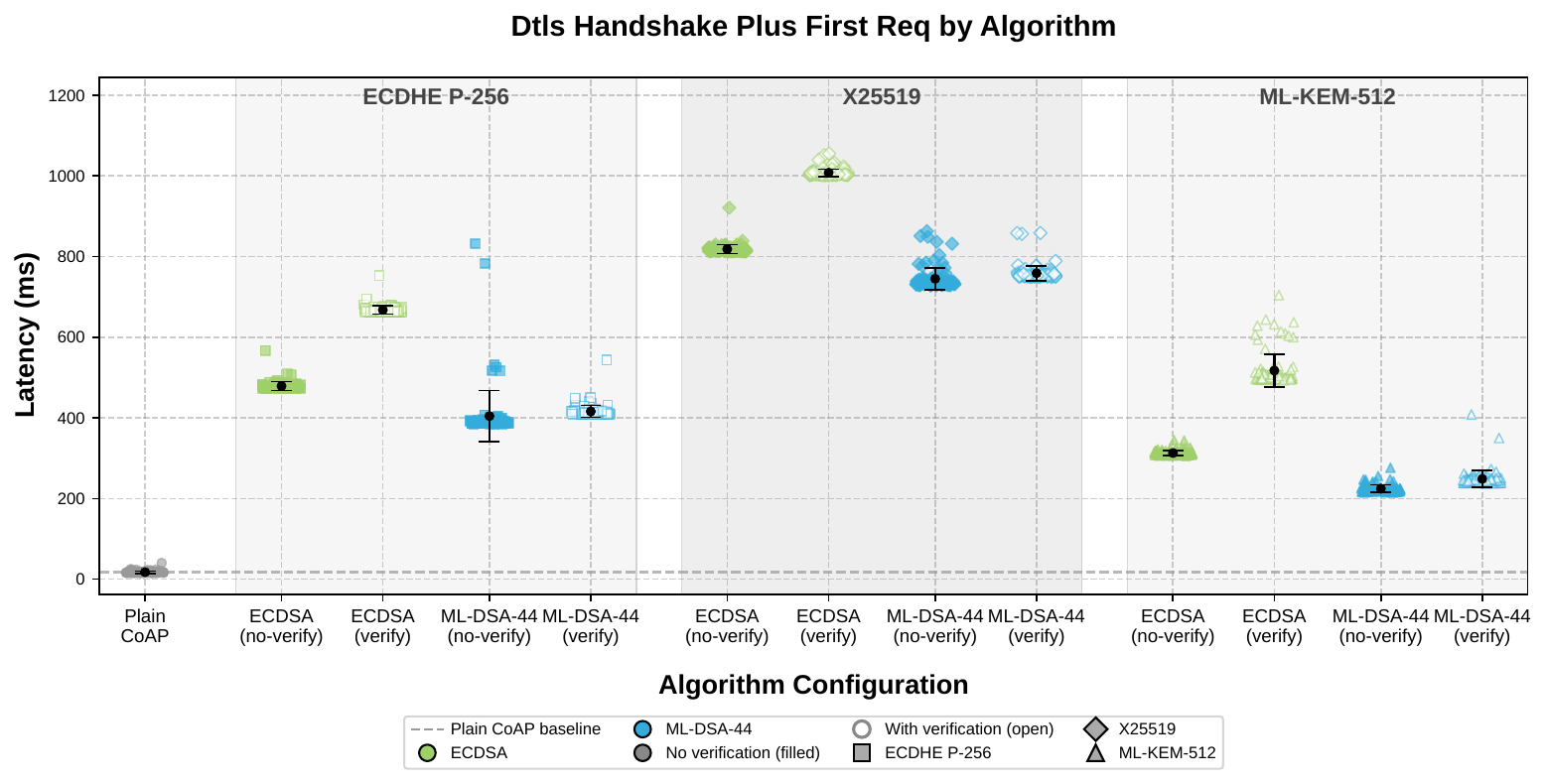}
  \caption{DTLS~1.3 handshake + first request latency distribution per algorithm combination. Each dot is one iteration (fresh session); black markers show mean $\pm$ 1\,std. Algorithms are grouped by key exchange (ECDHE~P-256, X25519, ML-KEM-512), with ECDSA (green) and ML-DSA-44 (blue) variants within each group. Filled markers: no certificate verification; open markers: verification enabled.}
  \label{fig:algo_scatter}
\end{figure}

Figure~\ref{fig:algo_scatter} shows the same result without averaging it away. The point clouds preserve the ranking ML-KEM-512 $<$ P-256 $<$ X25519 under both signature schemes and both verification modes. Within each key exchange, enabling verification simply shifts the cloud upward: by about \SI{194}{\milli\second} for ECDSA and only about \SI{17}{\milli\second} for ML-DSA-44.

Since local entropy pool operations are excluded from the Phase~2 timers and the symmetric record layer is algorithm-independent (${\sim}$\SI{8}{\milli\second}, Section~\ref{sec:evaluation}), the handshake measurements differ across algorithms only in the asymmetric key-exchange computation and signature operations performed on both client and server. The CertificateVerify message, a mandatory DTLS~1.3 component in which the server signs the handshake transcript and the client verifies this signature, is included in both modes; \texttt{WOLFSSL\_NO\_VERIFY} skips only the separate certificate chain validation. A handshake without any signing is impossible under certificate-based authentication because the CertificateVerify proves private-key possession.

Combining the network estimate with the directly measured verification overhead yields a three-component decomposition. The DTLS~1.3 handshake with HelloRetryRequest involves three network round trips (ClientHello$\to$HRR, ClientHello2$\to$Server\-Hello\allowbreak{}+\allowbreak{}Finished, Finished\allowbreak{}+\allowbreak{}request$\to$response). Using the plain CoAP baseline (\SI{17.6}{\milli\second} mean) as the per-round-trip cost, the estimated network overhead is $3 \times 17.6 \approx \SI{52.7}{\milli\second}$. Because the plain CoAP round trip includes server-side proxy and HTTP~API processing that applies only to the third flight (the first two flights are pure DTLS messages without application-layer overhead), this estimate slightly overestimates the network component. Accordingly, the residuals below should be interpreted as conservative lower-bound estimates of the combined client\,+\,server key exchange and signing time, not as directly measured computation times. Subtracting the estimated network component from each non-verify handshake mean yields lower-bound residuals of $172$\,ms for ML-KEM-512~+~ML-DSA-44, $260$\,ms for ML-KEM-512~+~ECDSA, $352$\,ms for P-256~+~ML-DSA-44, $426$\,ms for P-256~+~ECDSA, $692$\,ms for X25519~+~ML-DSA-44, and $766$\,ms for X25519~+~ECDSA.

\begin{figure}[H]
  \centering
  \includegraphics[width=0.95\linewidth]{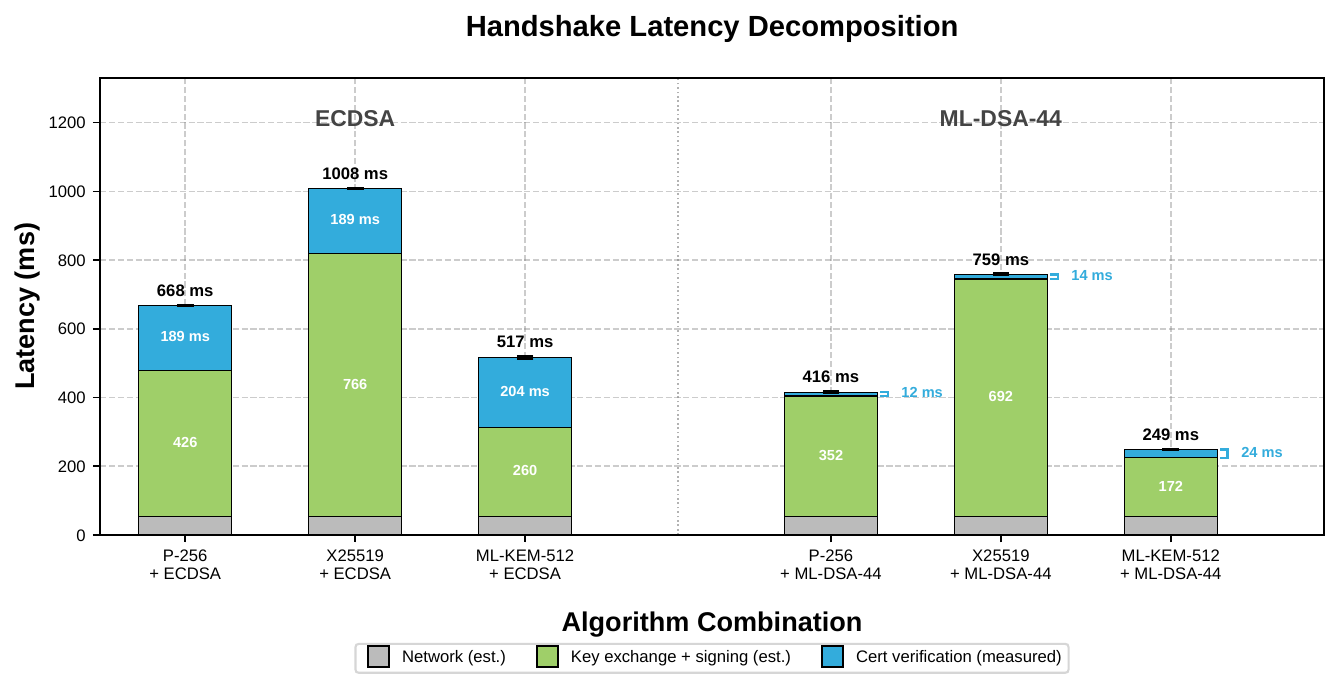}
  \caption{Decomposition of DTLS~1.3 handshake latency (with verification enabled) into three components: estimated network overhead (grey; 3 round trips $\times$ plain CoAP mean), lower-bound residual estimates for key exchange and CertificateVerify signing computation (green; non-verify residual after subtracting the estimated network component), and directly measured client-side certificate chain verification (blue; verify$-$non-verify mean). ECDSA group (left) and ML-DSA-44 group (right). Green segment labels show the lower-bound residual; blue annotations show the verification cost; totals above each bar give the full handshake latency, all in milliseconds.}
  \label{fig:decomposition}
\end{figure}

The third component, client-side certificate chain verification, is obtained directly as the difference between verify and non-verify handshake means (pooled: \SI{194.2 \pm 1.6}{\milli\second} for ECDSA and \SI{16.5 \pm 2.6}{\milli\second} for ML-DSA-44), requiring no model assumptions. Unlike the network/computation split, this verification overhead is an exact arithmetic difference. Using the lower-bound residual estimates above, the computational advantage of ML-KEM-512~+~ML-DSA-44 over P-256~+~ECDSA reaches 59.7\% without verification and 68.1\% with verification, substantially larger than the end-to-end improvement, because the fixed network overhead compresses relative differences. Figure~\ref{fig:decomposition} visualises this three-component decomposition for all six algorithm combinations, showing that key exchange and signing constitute the dominant cost component across all configurations, far exceeding both the fixed network overhead and the verification cost.

\section{Discussion}
\label{sec:discussion}
The dual-access server architecture fits two distinct use cases. HTTPS clients can consume entropy through the direct OpenSSL path, while constrained devices use the CoAP API and fold the received bytes into a local pool.

The main practical result is that the expensive part is still affordable. With 100 handshake traces per configuration, DTLS~1.3 setup ranges from \SI{224.6}{\milli\second} for ML-KEM-512~+~ML-DSA-44 without verification to \SI{1008.0}{\milli\second} for X25519~+~ECDSA with verification. That cost is paid once per session. Afterward, CoAP requests settle near \SI{24}{\milli\second} regardless of the asymmetric algorithm choice. The BLAKE2s pool stays far below the timing noise floor of the network, adding less than \SI{0.1}{\milli\second} for injection plus extraction.

In operational terms, an established DTLS session sustains about 41.5 entropy requests per second, given the \SI{24.1}{\milli\second} mean RTT. At 8 bytes per response, that is \SI{332.4}{\byte\per\second}, enough to refill the \SI{512}{\byte} pool in about \SI{1.54}{\second}. If the session is lost, ML-KEM-512~+~ML-DSA-44 with verification restores it in under \SI{250}{\milli\second}. That is fast enough for duty-cycled nodes that wake up, top up their entropy pool, and go back to sleep.

ML-KEM-512 is faster than ECDHE~P-256 here for a simple computational reason. Its encapsulation path is built from matrix-vector operations that map well to the Xtensa LX6, whereas ECDHE needs two elliptic-curve scalar multiplications: one for key generation and one for shared-secret derivation. ML-DSA-44 adds another gain because CertificateVerify signing and verification are both cheaper than their ECDSA counterparts on this platform (Section~\ref{sec:evaluation}). The tradeoff is memory, not speed. Relative to the classical baseline, the fully post-quantum configuration needs about \SI{30}{\kilo\byte} more flash and \SI{31}{\kilo\byte} more peak heap, but it cuts verified handshake latency by 63\%. This matches the direction reported by Sosnowski et al.~\cite{sosnowski2023performance} for TLS~1.3 on server-class hardware and shows that the same trend survives on a constrained microcontroller.

Getting PQC to run on the ESP32 was mostly an exercise in respecting memory reality. Zephyr with newlib exposes two separate heaps: a small sbrk-based heap of about \SI{22}{\kilo\byte}, used by \texttt{malloc()}, and a \SI{105}{\kilo\byte} system heap, used by \texttt{k\_malloc()}. wolfSSL defaults to \texttt{malloc()}, so early builds simply ran out of memory during handshake buffer allocation. Redirecting allocations through \texttt{XMALLOC\_OVERRIDE} fixed that. A second obstacle came from ML-KEM key shares exceeding the DTLS MTU. wolfSSL's ClientHello fragmentation path depends on server-side code and breaks with \texttt{NO\_WOLFSSL\_SERVER}. The solution was to use the standard HelloRetryRequest cookie flow (RFC~9147~\S5.1), which adds one round trip but drops the server-code dependency entirely; no cryptographic property changes, and the extra $\sim$\SI{18}{\milli\second} is already captured in all reported handshake latencies. From there, the rest was careful trimming: cap static buffers at the \SI{1400}{\byte} DTLS MTU instead of the default \SI{16384}{\byte}, remove unused cipher suites, and preserve just enough headroom for the worst case. The final verified ML-KEM-512~+~ML-DSA-44 build peaks at \SI{97}{\kilo\byte} out of the available \SI{105}{\kilo\byte} heap.

\section{Conclusions}
\label{sec:conclusions}
We built and evaluated a full QEaaS path for embedded systems under post-quantum security constraints. The server exposes both direct QRNG-backed entropy through OpenSSL and mixed entropy through the Linux pool. The client extends libcoap's Zephyr support~\cite{libcoap_zephyr_pr,libcoap_wolfssl_pr} and adds a BLAKE2s-based entropy pool for ESP32 microcontrollers.

The measurements on ESP32 hardware are straightforward. ML-KEM-512 completes a DTLS~1.3 handshake in \SI{313}{\milli\second} on average without certificate verification, 35\% faster than ECDHE~P-256. Pairing ML-KEM-512 with ML-DSA-44 reduces the mean to \SI{225}{\milli\second}. Full certificate verification adds only \SI{16.5 \pm 2.6}{\milli\second} on average for ML-DSA-44, compared with \SI{194.2 \pm 1.6}{\milli\second} for ECDSA. As a result, the verified ML-KEM-512~+~ML-DSA-44 configuration is 63\% faster than the verified ECDHE~P-256~+~ECDSA baseline, at the cost of about \SI{30}{\kilo\byte} additional flash and \SI{31}{\kilo\byte} additional peak heap. Once the session is open, CoAP entropy requests complete in \SI{24.1}{\milli\second} on average, and local pool handling contributes less than \SI{0.1}{\milli\second}. On this class of device, post-quantum transport is not merely viable; for the tested configurations, it is the faster option.

The next steps are clear: move entropy injection into Zephyr's OS-level work queue for automatic pool refill, repeat the study on ESP32-S3 hardware to test higher-security ML-KEM variants, and add an explicit energy analysis~\cite{tasopoulos2023energy} for battery-powered deployments.

\section*{Author Contributions}
Conceptualization, J.B.-R., F.A.M.\ and D.D.-S.; methodology, J.B.-R.; software, J.B.-R.\ and Y.M.G.-N.; validation, J.B.-R.; formal analysis, J.B.-R.; investigation, J.B.-R.; resources, D.D.-S.\ and C.G.-R.; data curation, J.B.-R.; writing---original draft preparation, J.B.-R.; writing---review and editing, F.A.M., D.D.-S., C.G.-R.\ and C.C.; visualization, J.B.-R.; supervision, F.A.M., D.D.-S.\ and C.C.; project administration, C.G.-R.\ and C.C.; funding acquisition, D.D.-S., C.G.-R.\ and C.C. All authors have read and agreed to the published version of the manuscript.

\section*{Funding}
This work was supported by the Spanish Government under the grant TED2021-130369B-C32 funded by MICIU/AEI/10.13039/501100011033 and by the ``European Union NextGenerationEU/PRTR'' and the grant PID2023-148716OB-C33 funded by MICIU/AEI/10.13039/501100011033.
This research has been also funded by project I-SHAPER: Internet-Service Hardening of Authentication, Confidentiality, Privacy, Enforcement, and Reliability, from the public invitation program for collaboration in the promotion of Strategic Cybersecurity Projects in Spain-PRTR-of the National Cybersecurity Institute of Spain (INCIBE) c114.23. In addition, this action has also been supported through the R\&D activities program with reference TEC-2024/COM-504 and the acronym RAMONES-CM granted by the Comunidad de Madrid, Spain, through the Directorate General for Research and Technological Innovation by Order 5696/2024.

\section*{Data Availability}
The server-side code is available at \url{https://github.com/qursa-uc3m/qeaas-server}. The client-side code is available at \url{https://github.com/qursa-uc3m/qeaas_esp32_client}. Benchmark data are included in the client repository.

\section*{Acknowledgments}
We thank Rodrigo Fern\'andez for the initial development of the CoAP-HTTP Proxy.

\section*{Conflicts of Interest}
The authors declare no conflicts of interest.

\printbibliography

\end{document}